\documentclass[manuscript,screen]{acmart}
\AtBeginDocument{%
  \providecommand\BibTeX{{%
    \normalfont B\kern-0.5em{\scshape i\kern-0.25em b}\kern-0.8em\TeX}}}

\begin{document}

\title{Addressing the Selection Bias in Voice Assistance: Training Voice Assistance Model in Python with Equal Data Selection\\
}
\author{Kashav Piya}
\email{kashavpiya19@augustana.edu}
\affiliation{%
  \institution{Augustana College}
  \streetaddress{639 38th St}
  \city{Rock Island}
  \state{Illinois}
  \country{USA}
  \postcode{61201}
}

\author{Srijal Shrestha}
\email{srijalshrestha18@augustana.edu}
\affiliation{%
  \institution{Augustana College}
  \streetaddress{639 38th St}
  \city{Rock Island}
  \state{Illinois}
  \country{USA}
  \postcode{61201}
}

\author{Cameran Frank}
\email{cameranfrank18@augustana.edu}
\affiliation{%
  \institution{Augustana College}
  \streetaddress{639 38th St}
  \city{Rock Island}
  \state{Illinois}
  \country{USA}
  \postcode{61201}
}

\author{Estephanos Jebessa}
\email{estephanosjebessa19@augustana.edu}
\affiliation{%
  \institution{Augustana College}
  \streetaddress{639 38th St}
  \city{Rock Island}
  \state{Illinois}
  \country{USA}
  \postcode{61201}
}

\author{Tauheed Khan Mohd}
\email{tauheedkhanmohd@augustana.edu}
\affiliation{%
  \institution{Augustana College}
  \streetaddress{639 38th St}
  \city{Rock Island}
  \state{Illinois}
  \country{USA}
  \postcode{61201}
}

\begin{abstract}In recent times, voice assistants have become a part of our day-to-day lives, allowing information retrieval by voice synthesis, voice recognition, and natural language processing. These voice assistants can be found in many modern-day devices such as Apple, Amazon, Google, and Samsung. This project is primarily focused on Virtual Assistance in Natural Language Processing. Natural Language Processing is a form of AI that helps machines understand people and create feedback loops. This project will use deep learning to create a Voice Recognizer and use Commonvoice and data collected from the local community for model training using Google Colaboratory. After recognizing a command, the AI assistant will be able to perform the most suitable actions and then give a response.

The motivation for this project comes from the race and gender bias that exists in many virtual assistants. The computer industry is primarily dominated by the male gender, and because of this, many of the products produced do not regard women. This bias has an impact on natural language processing. This project will be utilizing various open-source projects to implement machine learning algorithms and train the assistant algorithm to recognize different types of voices, accents, and dialects. Through this project, the goal to use voice data from underrepresented groups to build a voice assistant that can recognize voices regardless of gender, race, or accent. 

Increasing the representation of women in the computer industry is important for the future of the industry. By representing women in the initial study of voice assistants, it can be shown that females play a vital role in the development of this technology. In line with related work, this project will use first-hand data from the college population and middle-aged adults to train voice assistant to combat gender bias.

\end{abstract}


\maketitle

\section{Introduction}

The first-ever voice-activated consumer product was released to the public in 1922. It was known as “Radio Rex.” This product was a toy that had a doghouse with a dog inside it. When someone said “Rex” next to the dog house, the dog would jump out of the doghouse. This voice-activated toy was created even before modern computers existed.\cite{subhash2020artificial}

Since the development of that toy, there has been a considerable amount of development in voice recognition, natural language processing, and machine learning. A voice assistant, also known as an intelligent personal assistant or a connected speaker, is based on natural language speech recognition. Recently it has had a rise in popularity and has been marketed and used by Apple, Amazon, Google, and Samsung. Now voice assistants are widely found in most modern-day devices that a person would use. 

Voice assistants are multi-purposed one of their main purposes was for a search to be carried out using a voice command entered by the user as an input. They are also known to be used for information retrieval by voice synthesis. They use a variety of voice recognition techniques, language processing algorithms, and voice synthesis to listen to specific voice commands that may include wake words, tasks, and queries, and return relevant information or perform a specific function as requested by the user. These assistants can be software-based which allows them to be integrated into a wide range of devices such as laptops, mobile devices, and speakers, or can be specifically designed into a standalone device like Amazon Echo or Amazon Alexa Wall Clock. \cite{nasirian2017ai}

These voice assistants work like a charm and are quite fascinating, making one might ask themselves what goes on behind the hood of these amazing innovations or how do they work the way they do? 

To answer the above query, in short voice assistants use artificial intelligence and voice recognition to deliver the result that the user is looking for efficiently, and precisely. The user provides a command to the voice assistant that is called intent. Through voice recognition, these intentions can be understood by our virtual assistants. Here, voice recognition allows the speaker to speak into a device that takes the analog signal from the speaker and changes it into a digital signal which is then processed by the computer to match it with words or phrases and then recognize the command. Machine learning also has a huge part to play in this as the computer needs to be taught to recognize the speaker's words by feeding it a database of words and syllables in each language to match it with digital signals. This process is known as pattern recognition. Additionally, these devices gather a lot of information from the commands that they received previously to improve upon themselves using machine learning.\cite{kudina2021alexa}

There are multiple approaches to voice assistants, specifically two types:  task-oriented and knowledge-oriented. Most voice assistants these days can combine both the task-oriented as well as a knowledge-oriented workflow to complete all the tasks that a user may ask the voice assistant to carry out. A task-oriented approach will most likely ask something as simple as filling out a form, whereas a knowledge-based approach may include answering questions such as who the President of the United States of America is or finding out what engine is in a Ford F50 which is a technical specification of a product. \cite{chattaraman2019should}

The task-oriented approach/workflow is pretty much self-explanatory as it uses goals and tasks to achieve whatever the user wants or needs. This approach usually requires the voice assistant to use a different application such as time, weather, web browser, and music apps, to help complete its tasks. Some examples would be, asking a voice assistant to set a reminder to take medicine at 6 PM, playing music using Spotify, etc. This approach does not require the virtual voice assistant to search massive databases for knowledge. These tasks are often known as skills. And various assistants allow for different skills to be installed according to the user’s preferences. \cite{schmidt2018towards}

Whereas a knowledge-oriented approach/workflow requires the use of analytical data to complete the tasks and help the users to complete their tasks. \cite{bernaras1994problem} Unlike a task-oriented approach, this approach focuses on using online databases to get related information in addition to already recorded knowledge to help users to complete tasks. An example of a knowledge-based approach would be if a user asked for a question that would require searching the internet such as what is the capital of the state of Illinois or who invented the telephone?
	
Furthermore, there are two types of artificial intelligence (AI); In general, there is a weak AI and there is a strong AI.\cite{bringsjord2003artificial} There are many types of machines such as Siri, Alexa, Cortana, and Bixby that can only perform certain tasks that have been defined by the user while making the AI. These types of AIs are called Weak AI. And some machines or systems have a mind of their own and can make decisions or take actions on their own without human interference. These types of machines are called Strong AI. After learning the differences between strong and weak AI, the voice assistant this project is opting for is an example of weak AI. 

This projects Virtual Voice Assistant will include a variety of features such as greeting the users, fetching information about a person, an object, or anything else in general from the internet, providing the time, opening web browsers, playing music, and so on. It might also include additional features such as opening the web camera to take pictures, forecasting the weather, logging off from your personal computer, telling you a joke, and many other features.

The field of Virtual Assistance has many avenues to consider from providing help in technology to connecting people through the usage of technology. When studying what exactly a Virtual Assistant is, the field that was decided on was Virtual Assistance in natural language processing, which means the technology can understand people more accurately. Natural Language Processing is a form of AI that gives machines the ability to not just read but to understand and interpret human language. With NLP, machines can make sense of the written or spoken text and perform tasks including speech recognition, sentiment analysis, and automatic text summarization \cite{hendrix1978developing}. Therefore, not only does natural language processing help humans it also helps with machine learning, in the sense that NLP will continue to provide more data to better that analysis of speech and create a feedback loop.
	
The English language is an extremely hard language to understand and speak, especially if English is not your first language. Not only is the usage of English vernacular hard to comprehend and execute, but there is also a form of a language barrier in the different dialects that people possess. A person's dialect can be a communication inhibitor in many languages, not just in English, but in this project, the focus is on the English language  \cite{wold2006difficulties}.
	
For this project, synthetic voices were originally used instead of human voices, in which data was collected. A synthetic voice is a pre-recorded voice produced through text to speech whereas the human voice is pre-recorded. ‘Its use involves recording, in advance, a text read aloud by a human being.’ The usage of a synthetic voice will give flexibility as they have a high capacity in reading textual context and can generate voice constantly. But “there is a disadvantage in expressing social signs as it cannot express emotions, intentions, and attitudes through modulation of the voice” \cite{beller2007content}.
	
The computer industry is primarily dominated by the male gender and because of this extreme one-sided representation in the field a multitude of the products that are produced do not regard women. One of the fields that this bias impacts is natural language processing. Because of the lack of females in the industry, the identification percentage for female voices is lower than that of male voices, therefore resulting in the analysis and research of this topic of selection bias of voice assistance \cite{caliskan2021}. Data augmentation by controlling the gender attribute is an effective technique in mitigating gender bias in NLP processes.

\section{Related Work}
\subsection{How Does Voice Assistant Work? }

\begin{figure}[htp]
    \centering
    \includegraphics[scale = 0.75]{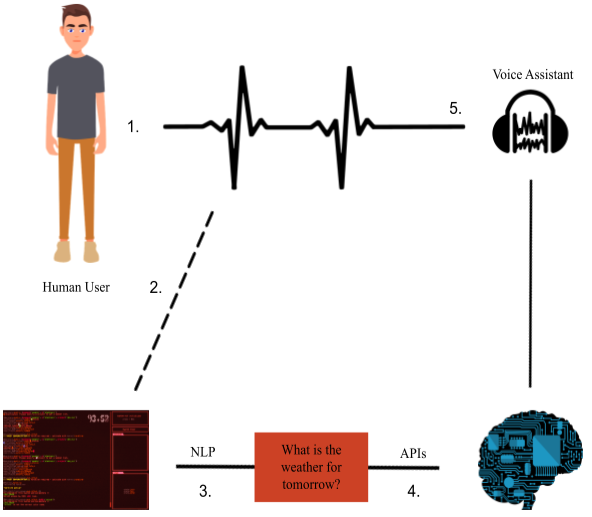}
    \caption{How Does Voice Assistant Work?}
    \label{fig:VoiceAssistant}
\end{figure}

Voice assistance has now been defined is, but how does it work? A voice assistant uses speech recognition along with other identification of speech components to help the machine process the voice. Then, the speech is rendered into its textual representation based on extracted patterns. Following that, this program isolates the most important words or the action also known as anticipated intent. If the intent is not clear, the voice assistant is programmed to ask more questions. It then retrieves information by API calls to access the relevant knowledge base. Finally, it relays the information back to the human user through text to speech or fulfills the necessary action. Voice assistants rely on Natural Language Processing and other machine learning algorithms to perform the best and overcome the challenges that it faces. \cite{chowdhary2020natural}

\subsection{Speech Recognition}

How does a voice assistant understand what a person is talking about? These voice assistants use fast decoding algorithms that allow real-time continuous speech recognition systems to provide instant responses.\cite{reddy1976speech} Speech Recognition has been a part of our day-to-day life for more than 10 years now as it seems to become more and more advanced. The beauty of speech-to-text models goes unnoticed in this process as the machines make it look so seamless. Most speech recognition algorithms use a mix of Natural Language Processing and Deep Learning techniques to parse through the user query, get an appropriate response and present it back to the user in whichever form the user desires. All the big companies such as Google's Assistant, Amazon's Alexa, Apple's Siri, and others use the same techniques but just with some different variations.\cite{meyer2013speech} 

\subsubsection{Signal Processing for Speech Recognition}

Audio signals are any object that vibrates to produce sound waves. When an object vibrates, the air molecules oscillate to and from their rest position and transmits its energy to the neighboring molecules. This results in the transmission of energy from one molecule to another which in turn produces a sound wave. \cite{laroche2002time}

There are a few terms that should be familiar when talking about sound and signal processing. 
\begin{itemize}
    \item Amplitude: Amplitude refers to the maximum displacement of the air molecules from the rest position
    \item Crest and Trough: The crest is the highest point in the wave whereas trough is the lowest point
    \item Wavelength: The distance between two successive crests or troughs is known as a wavelength
    \item Cycle: Every audio signal traverses in the form of cycles. One complete upward movement and downward movement of the signal form a cycle
    \item Frequency: Frequency refers to how fast a signal is changing over a period
\end{itemize}

Additionally, there are two different types of signals: Digital and Analog Signal. For digital signal is a discrete representation of a signal over a period. Here, the finite number of samples exists between any two-time intervals whereas the analog signal is a continuous representation of a signal over a period which implies that there will be an infinite number of samples between any two given time intervals. \cite{kim1999auditory}

For this project, audio signals are needed for the Voice Assistant, so there is a question about how it stores a signal that has an infinite number of samples since they are analog signals. This project will incorporate changing the memory-hogging analog signal using complex techniques to convert it into digital signals to make it more convenient to work with them.

To convert a signal from analog to digital, a technique called sampling the signal should be used which selects a certain number of samples per second from the analog signal which makes storing and processing the signal memory efficient. In analog to digital sampling "an input signal is converted from some continuously varying physical value (e.g. pressure in air, or frequency or wavelength of light), by some electro-mechanical device into a continuously varying electrical signal." \cite{crowcroft1998analog}

\subsubsection{Feature Extraction Techniques for Audio signals}

For this projects model to use the audio signals, the features must be extracted from the audio such as time domain and frequency domain. The audio signal is representing by the amplitude as a function of time. The features here are the amplitudes which are recorded at different time intervals. On the other hand, in the frequency domain, the audio signal is represents amplitude as a function of frequency where the features are amplitudes that are recorded at different frequencies.

\subsection{Papers Regarding this Topic}
In machine learning when one trains their model, there is always a chance for problems when applying the model to the population. This problem arises in big data when the population is too large for machines with limited computability. To deal with this problem of big data, a sample population can be used. A sample population refers to a subset of the population that represents the whole population. \cite{israel1992determining} This project will apply machine learning to samples population. But there is a chance for selection bias, “selection bias is a systematic error that results in differences between a study population and a target population; selection bias primarily affects the external validity of the results of a study”. This means that the results thought to be true for the sample population may not be true for the actual population. Depending on the population sample, if there are too many true results it will just show true no matter what. Sample population can technically be wrong from the population just because of the sample collected. 

Likewise, voice assistants use sample data to analyze a user’s speech. “Currently speech recognition has significant race and gender biases. It is just another form of AI that performs worse for women and non-white people. Currently, it is designed to understand white male voices well” \cite{bajorek2019voice}. This is a significant problem because these days voice assistants are a crucial part of people’s lives from setting alarms to transportation. Low accuracy in voice recognition would mean severe consequences in people’s life \cite{bajorek2019voice}.
In the paper Empirical Analysis of Bias in Voice-based Personal Assistants, they try to check the accuracy of relevant voice assistants such as Google Assistant and Siri. They look at the different accents for the Brazilian Portuguese, and how accuracy was off for certain accents than other ones. They also found variation in the quality of recognition based on gender. \cite{lima2019empirical}

So, taking the two paper’s ideas, to tackle this problem and move forward in this topic, data was gathered and training our model not only on prevalent male voice data sets but also gathering voices manually and from Common-Voice data sets by Mozilla for female and minority voices. This allows us to create samples proportionally to demographic indicators of the country. The process for machine learning will be as follows:
\begin{itemize}
    \item Filter the words that the user says
    \item Digitize the user’s speech into a format that the machine can read
    \item Analyze the user’s speech for meaning
    \item Decide what the user needs based on previous input and algorithms” 
\end{itemize}
 
As talked above, data sets for voice recognition must be sampled so that each voice has an equal probability to be included in the sample. This allows for less racial and gender bias for the models to work with. So, this will result in everyone having their voice heard. 

Additionally the paper “Dangerous Skills: Understanding and Mitigating Security Risks of Voice-Controlled Third-Party Functions on Virtual Personal Assistant Systems” \cite{zhang2019dangerous} and “Your Voice Assistant is Mine: How to Abuse Speakers to Steal Information and Control Your Phone” talks about the development of voice assistant in speech recognition and IoTs and with its development comes more vulnerabilities. They talk about voice-based remote attacks and permission bypassing. These problems are extremely dangerous and can be misused to expose people’s information. So, these papers advise incorporating the ideas of not allowing zero permission and context-aware information collection and analysis features to the voice assistant. Not allowing those ideas will allow focusing on forcing restrictions on specific operations that a particular process can perform \cite{diao2014your}.

\section{Experimental Setup}
The data used is partially from Common-Voice, which is an online open source of voice recorders in multiple languages. However, since this project is specifically for the English language the data set collected was English. However, this data set, as expected has more male voices than female voices. Therefore, to even out the distribution between the voices represented the rest of the data was collected via outreach into the Augustana College community, contacting over 200 female students (ages ranging from 18 - 24 years old) and receiving about 65 voices samples with 4-7 minutes of voice recording from each sample. The data set collected was then doubled, converted, and combined into .wav files. 

After the data collection and manipulated, it was then applied to a gender recognition model in order to numerically identify via the frequency of each voice; according to ASHA (American Speech Language and Hearing Association) the average range for an adult woman is 165 to 255 Hz and the average range for adult males is 85 to 155 Hz \cite{watson2019unheard}. There is inherent technical problem down to the fact that females generally have higher pitched voices. Female voices tend to be quieter and sound more “breathy”. Female more easily masked by noise, like a fan or traffic in the background, which makes it harder for speech recognition systems.

In order to run the data through training model for speech recognition. The full data set from Common-Voice was not needed in fact there only segments of the a statement is used, therefore it was necessary to parse the data into smaller form.

\section{Framework}
For Machine Learning, Python is known to be the most common language used due to its versatility, readability, and abundant packages. So, for this project, it will be using Python 3.10 as the main programming language. This project will be using the following python packages or libraries:
\begin{itemize}
    \item Speech Recognition (Will only be used initially to create and test the basic functionalities of the project and will be later replaced by this projects own model): It helps understand what the human is saying and converts the speech into text.
    \item Pyttsx3: It is a simple text to speech conversion library in python which will be used to give this projects Voice Assistant a voice.
    \item Wikipedia: This package is a Python that extracts information and data from Wikipedia which is a multilingual online encyclopedia used by many people.
    \item Capture: It helps to capture images from your camera.
    \item Datetime: It is an inbuilt module in python that works with date and time.
    \item Os: It provides functions to interact with the operating system.
    \item Web Browser: It is an in-build package from Python that allows you to extract information from the web.
    \item JSON: It is a module that helps to store and exchange data.
    \item PyJokes: It gives the user a random joke.
    \item Pyaudio: It allows python to play and record audio between different platforms.
    \item Pywhatkit: It can access YouTube in order to play a video.
    \item Librosa: It is a package for analysing sound, audio and music.
    \item Soundfile: This package can help read and write sound files.
    \item Numpy: Numpy provides a powerful N-dimensional array object, sophisticated functions, tolls for integrating C/C++ and Fortran code, useful linear algebra, Fourier transform, and random number capabilities, and much more. 
    \item BeautifulSoup: Beautiful Soup is a Python library for pulling data out of HTML and XML files.
    \item Pandas: pandas is a fast, powerful, flexible and easy to use open source data analysis and manipulation tool, built on top of the Python programming language.
\end{itemize}

This project will be using all these modules and packages to create basic functionalities of the Voice Assistant. For example, this project will use the Wikipedia package to get information regarding a person or a company. This function is used to grab the first few sentences from its corresponding Wikipedia page which is generally the broad introduction to the subject.  

Then this project will have a Voice Recognizer which will replace the existing Speech Recognition module as well as the Google API. The program uses deep learning to create this projects voice recognition using PyTorch. The data is from open sources Mozilla Common-Voice as well as a few more from Kaggle to train this projects model and add voices from volunteers to help balance the data to reduce the biases while training for the wake word as well as the general voice recognition itself. This project utilizes Google Colaboratory to train this projects model. Finally, if time permits, the goal is to build a simple User Interface either with Flask or with other python libraries itself for a visual element to the program.

\section{Methods}
As stated in the experimental setup, the Common-Voice data set planned to be used for the training model has recognized 46\% male voices and 16\% female voices. Therefore, the goal of this project is to come up with an effective solution to this disparity.

When training a speech recognition model, different types of voice data can be used. Depending on the type of interaction one’s looking to build, and how robust that interaction should be, different types of voice data might be required. Although there are several easily available sources of speech data, such as public speech corpora or pre-packaged data sets, it is almost always necessary to cooperate with a data services provider to collect your own speech data, either remotely or in person. When you gather one’s own data, it is easy to tailor the speech data set to include variables such as language, speaker demographics, audio requirements, and collection size. \cite{summalinguae2022}

The Bureau of Labor Statistics (BLS) projects computer science research jobs will grow 19 percent by 2026. However, in the United States, the percentage of women that receive a bachelor’s degree in Computer Science is still only 18 percent. There is currently a high demand for computer scientists in the professional industry but despite this fact, this industry remains male-dominated, in the United States. For example, in this year's Senior Inquire class for Computer Science, there is a 1:6 female to male ratio \cite{zilberman2021computer}. 
	
Computer technology first emerged during World War II and continuing into the 1960s, women made up most of the computing workforce. However, by 1970 women only accounted for about  14 percent of bachelor's in computer science. In 1984 that number rose to 37 percent. The percentage of women in computer science has since declined to 18 percent. It was around the same time personal computers started showing up in homes. According to NPR, personal computers were marketed almost exclusively to men and families were more likely to buy computers for boys than girls. 

Computers are now commonplace in both classrooms and on individuals as personal assistance. It is hard, however, to explain the exact reason why females are not as present in this major. There are organizations now that are researching and improving ways to increase the potential of more females in the computer science major. It is said that one of the reasons why women tend to trend away from the computer science field is because of marketing of the industry in the past tailored to those with the geek persona and the social innuendos of what being a geek used to mean. \cite{carter2006students}

One of the reasons why females should be represented in the computer industry is increasing the inclusion of women is a sound business strategy. “A study by Deloitte found that women's choices account for up to 85 percent of buying decisions nationwide, and that diversity drives innovation. Though it is still commonplace to find boards and project teams without a female member, the integration of female perspectives will naturally lead to higher revenues and a better understanding of consumer marketplaces.” \cite{computerscience2022}

Therefore, the purpose of this project is to strive to increase the attractiveness of the potential that computer science avenues provide for women, by increasing the representation of females in the initial study of voice assistants. If females can realize that they possess a more essential in the initial makeup that products that are generated there is a possibility that females will be more inclined into joining the computer science industry, by showing them that they do in fact play an essential role in the make of technology.

Speech recognition data can be classified into three categories:
\begin{itemize}
    \item Controlled: Scripted speech data
    \item Semi-controlled: Scenario-based speech data
    \item Natural: Unscripted or conversational speech data
\end{itemize}
For this project, the primary type of voice data used is Semi-controlled data and natural unscripted conversational speech data. For semi-controlled data, When developers need a natural sampling of different ways to ask for the same thing or a greater diversity of command intentions, scenario-based voice data is collected (i.e. asking for different things). As a result, scenario-based speech data adds variety to what is said as well as how it is said.

On the other hand, The most "natural" kind of speech is unscripted or conversational speech data, which is a recording of a conversation between two or more speakers. Unscripted speech data, like spontaneous speech, occurs in a variety of formats in the real world. For example, this information could be captured in the form of phone conversations or recordings of individuals conversing in a busy room. If a developer is looking for conversational data on a given topic (for example, music), two speakers might be asked to conduct a conversation about it.

\begin{figure}[h]
    \centering
    \includegraphics[scale = 0.75]{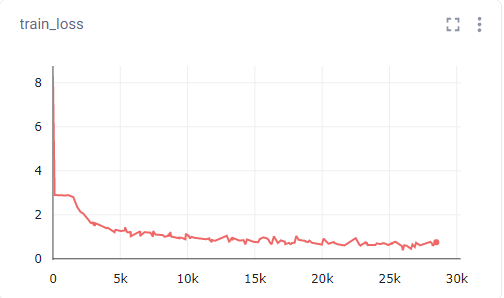}
    \caption{Train-Loss Graph}
    \label{fig:VoiceAssistant}
\end{figure}
\begin{figure}[h]
    \centering
    \includegraphics[scale = 0.75]{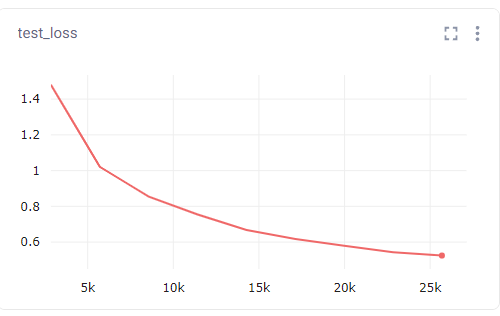}
    \caption{Test-Loss Graph}
    \label{fig:VoiceAssistant}
\end{figure}
\begin{figure}[h]
    \centering
    \includegraphics[scale = 0.75]{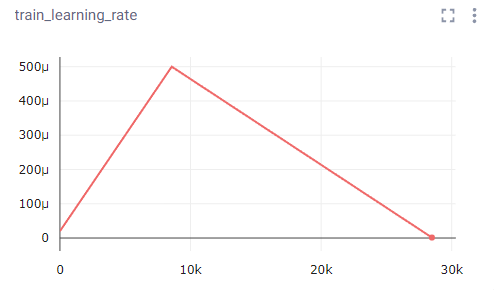}
    \caption{Training Learning Graph}
    \label{fig:VoiceAssistant}
\end{figure}

To compensate for this gender disparity, data augmentation will be done on the voice data-set in order to  artificially increase the diversity of the data-set and to increase the data-set size. This technique is performed by changing the pitch, speed, injecting noise, and/or adding reverb to the audio data.

The model was trained with about 30,000 separate lines of data and was trained over 7 epochs with a learning rate of 5e-1 and batch sizes of 15.

The train loss graph shows that the speech recognition model has more than enough data to train the model and is using more than enough data making the model almost over-fit which is also supported by the test loss graph.

As, shown by the train loss graph, the test loss graph also gives the same conclusion that the speech recognition model might be slightly over fitting since the train loss is very low, compared the test loss which is much higher than the train loss itself. However, the test loss is no too huge which makes the model usable. 

Finally, after half of the training, the learning rate started slowing down as the model kept getting more and more training with each epochs.

\section{Results}
The goal of this project is to create a voice-generated virtual assistant using python that can work similarly to the modern popular voice assistants such as Siri, Alexa, or Bixby. This projects personal AI voice assistant can understand voice commands using speech recognition in Python and will be able to perform multiple tasks, using the pre-programmed functions built into it as well as training data. 

At the end of the project, this project will have AI voice assistant to be able to recognize voices and their commands and perform the most suitable actions. This process of voice recognition is done by breaking down audio into individual sounds, then converting them into a digital format that will be using Machine Learning algorithms and models to find the word for that sound. The words will then be used by the voice assistant and see if anything related to it has been pre-programmed and if not, it will be able to perform a most suitable action. The assistant will be speech-enabled, so after recognizing a statement or a question, it will call the necessary function to execute the task then give a response based on the algorithm. 

However, the project will incorporate some features such as sending text messages, sending emails, opening songs, predicting the time, telling jokes, surfing the internet for information, forecasting weather, and many more. The final project will not be limited to the features listed above and will possess more as well as having a clearer vision of what this projects personal voice AI can do.

\begin{figure}[h]
    \centering
    \includegraphics[scale = 0.75]{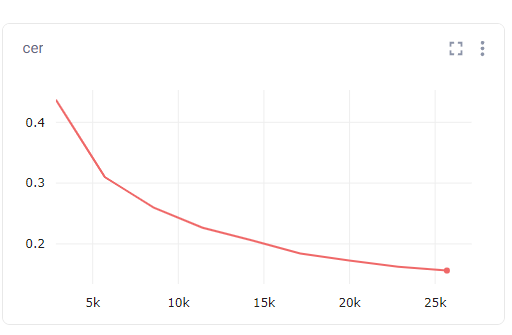}
    \caption{Cer graph}
    \label{fig:VoiceAssistant}
\end{figure}
\begin{figure}[h]
    \centering
    \includegraphics[scale = 0.75]{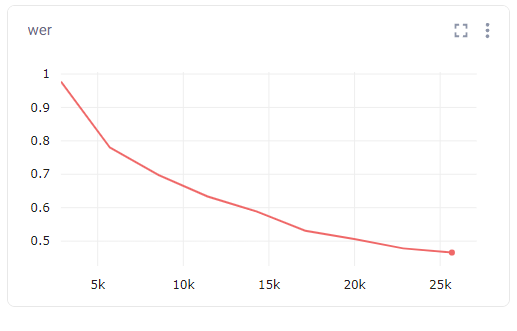}
    \caption{Wer graph}
    \label{fig:VoiceAssistant}
\end{figure}

For the model that was trained, the Wer(Word Error Rate) was reduced to about 50\% which is not so great when compared to other models made by large scale companies, such as Google's 4.9\% WER and Microsoft's 5.1\%. Also from the trained model the Cer(Character Error Rate) to about 20\% which means one out of every 5 characters were predicted incorrectly which is not the best.

\begin{figure}[h]
    \centering
    \includegraphics[scale = 0.40]{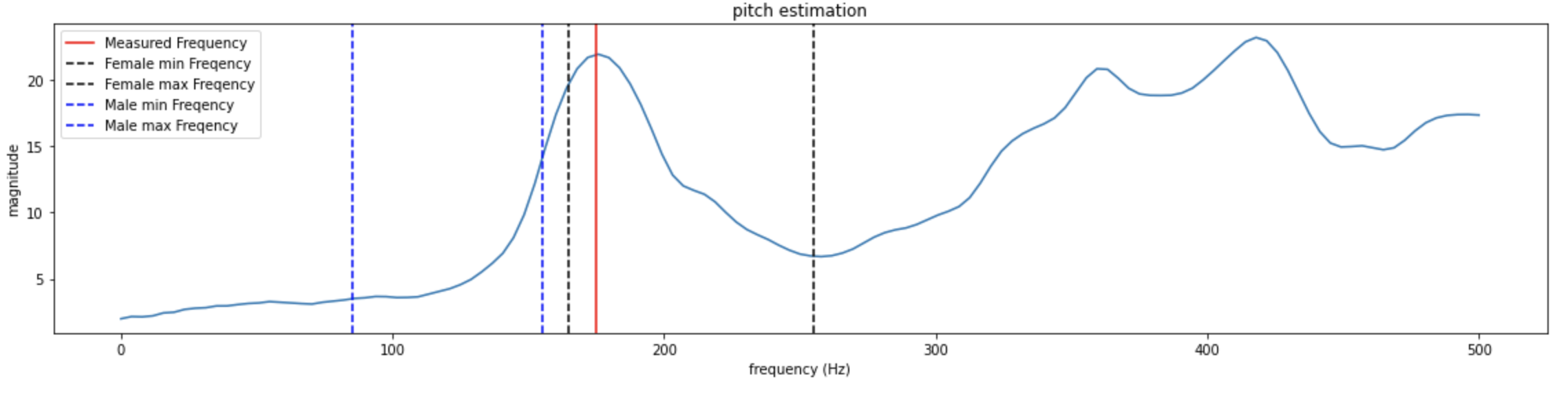}
    \caption{Pitch Estimator graph}
    \label{fig:VoiceAssistant}
\end{figure}

Part of this project was to see if the computer could correctly analyze whether a voice is female. Using the data set collected from Augustana 
College where every of the 62 voice samples are females. Using the frequency ranges mentioned earlier in the paper reported by ASHA as the boundaries for determination the computer categorized that 34 out the 62 voice samples are female voices meaning that the computer correctly estimated about 55\% of the samples. Figure 7 is a sample of one of the correctly categorized voices samples showing the frequency boundary ranges for both male and female voices with the red line be the voice samples calculated frequency. This results proves the hypothesis that computers simple have a hard time distinguishing female voice samples regardless of the frequency ranges.

One of the biggest challenges is training the assistant to recognize different types of voices, accents, and dialects. To counteract this problem by researching effective machine learning algorithms to implement to train the necessary voice data. This project will utilize various open-source projects to complete the voice recognizer which then can incorporate into this project own voice assistant.

One of the primary goals for this project is to build and train a model that utilizes more voice data from underrepresented groups, as there is a huge gender and race bias in most virtual assistants. This statement above can be implied because the training data used in some of the original voice assistants consists mainly of Caucasian males and Asian males. The female group as well as people from other backgrounds that might have different accents or dialects are extremely underrepresented. These models store the public data to try and improve the accuracy of the voice recognizer in their voice assistants, but will try to incorporate these data from the beginning. In the end, the voice assistant will be able to recognize voices regardless of their gender, race, or accent.

Another goal is the addition of a two factor authentication for added security to the user and their data. The two factor authentication will be set to access the history of all the commands provided by the user. The collection of data is done for the functionality of the virtual assistant. As well as allowing the user to keep track of the commands they have used. The goal is to have this authentication app ready for the final project, however this is now part of a future endeavor. A downside is that the voice recognition may not be viable, because it requires each user to train their voice specifically to the virtual assistant. This process will also require a lot more storage of data and processing on the programmers part.

This project demonstrates that gender and race biases exist in a lot of virtual assistants and tries to fix this problem by training more voice data from underrepresented groups. Males, especially white and Asian men, are disproportionately over represented in Computer Science education and careers. Because of the male-dominated employment milieu, many women abandon Computer Science careers. Researchers have found many hurdles to women in Computer Science courses at the academic level, although efforts to address these issues have differed depending on geographic region or educational level.

To combat this over-representation, gender depiction standards, as expressed through look, name, and voice, as well as a review procedure, are critical changes that must be introduced. Users are predisposed to draw gender from a voice, and often ascribe one to voices that are supposed to be neutral, as the makers of the gender-neutral voice assistant Q discovered \cite{robison2020}.

\section{Discussion}

Virtual Assistants take in voice commands and process that data into commands or useful information. In this age, many developed countries can already be considered as aged societies \cite{yamazaki2012home}. This causes lots of changes such as a discrepancy in the proportion of seniors with the rest of the population. So, assistance technology such as a virtual assistant can help with the problems associated with it. In the article “Home-Assistant Robot for an Aging Society” they provide information on how assistance technology helps in labor support, healthy lifestyle support, and household and care support. In consideration to virtual assistants, they give a list of activities done by an IRT home assistant robot that models 3D space in a vector space and uses voice recognition with reinforcement learning to do several tasks:
\begin{itemize}
    \item Performing Chores in a Home Environment
    \item Deformable Object Detection through image-based learning
    \item Geometrical Object Modeling and Its Application to Position Estimation
    \item Manipulation of Daily Tools and Appliances
    \item Integration of Basic Tasks Into Sequential Behavior
    \item Failure Detection and Recovery
\end{itemize}

The IRT home assistant robot works with voice commands to learn tasks. One can guide it to perform physical activities such as carrying items, pushing objects, collecting items, and sweeping. For example, it can take images of clothing and with vector data, learn about wrinkles in clothing to pick up clothes. The robot with more data can do the task without any commands. Likewise, it uses environment recognition to do similar tasks at several locations of the house using commands. The implementation of this assistance technology can increase the ease and productivity of performing home activities. These assistants not only help old people but also physically disabled and blind people as it does not require learning and typing where they can just speak and ask questions to the assistant.

In the expected results, this project discussed some of the outcomes of this projects virtual assistant. Again, it is still in the beginning phase, however, the main goal is to make it so that the assistant can do tasks like Siri or Alexa. “Virtual assistants are regularly used to make online interfaces more user friendly. It also generates positive responses from Internet users leading to a more interpersonal shopping experience, greater pleasure, and customer flow” \cite{holzwarth2006influence}.

In line with the related work, this project will be aiming to tackle the problem of the under-representation of women's voices and dialects for training. This project will be collecting first-hand data through participatory surveys and training our model based on that. Then this project will use data from online databases to further train our voice assistant. By doing this, this project can avoid the gender bias present in voice assistants. To confront the bias for dialects, this project will also be collecting voice command data from a diverse audience. This helps create a proportional sample that will help represent the college population and middle-aged adults that is the target.

Currently the project is not a proper application. It was presented with a GUI (Graphic User Interface) to access this project. In the future the goal is to implement an startup application with a better GUI that will run as soon as you turn your computer on. Since the application at the moment is only a GUI that will only receive commands there are no security risks. The program was also implemented so that the voice assistant will only listen when you press the listen button. In the future the plan is to make it so you can use your voice in order to activate it from the start by using a wake word. Doing so will make it easier to use but the goal is for the voice assistant to not collect sensitive information when the wake word is used by accident. To counter that the plan is to add a mute feature for the assistant to not allow it to listen. The future plan is also to add an indicator for when the voice assistant is actively listening as well as keeping logs of when listening occur ed and a text alert through a mobile device. But when developing the full application there will be a lot of potential security risks to consider. The plan to add several measures to manage the risks. One of the ways would be to personalize the voice assistant to one account unless one links another one. The future plan also includes adding a feature to ask for two-factor authentication in order to access the device use and history. This will help mitigate potential risks of identity theft or user impersonation. It will also add an extra layer of security to protect sensitive data from theft. The two-factor authorization that is used will be a pattern and/or pass-code and a token, or biometric data and a token. For the token, it will have a token generator or an authentication app. For the pattern and/or pass-code, this project will make sure that people can input it through their phones. As for biometric data, this project will add a part to the voice assistant so that it will learn the user's tone and pitch if allowed. 

As for the future of the project, the voice assistant will keep on building up with more features aimed for making the day to day activities easier for the user. Our research on the speech recognition will continue to grow as well as trying on new alternatives such as adding more layers to our neural network, using audio books to train our model, and using various other architectures available such as ctcdecode which can only be used in Linux environment which will improve our word error rate dramatically. The end goal of this project is ambitious considering that there are many speech recognition systems out already that perform very well, and have resources that cannot be compared to ours, but there will be improvements in the speech recognition system with more training and additional libraries.

\pagebreak

\bibliographystyle{ieeetr}
\bibliography{references}

\vspace{12pt}

\end{document}